\documentstyle[12pt]{article}
\author{A. Jadczyk \\
 Research Institute for Mathematical
Physics\\ Kyoto University, Kyoto 606, Japan \\and\\
Institute of
Theoretical Physics, University of Wroc{\l }aw, \\ Pl. Maxa Borna 9, PL 50
204 Wroc{\l }aw, Poland\thanks{Permanent address}}
\title{Particle Tracks, Events and Quantum Theory}
\date{ }
\def\complex{{\kern .1em {\raise .47ex
\hbox {$\scriptscriptstyle |$}} \kern -.4em {\rm C}}}
\def\be{\begin{equation}}
\def\ee{\end{equation}}
\def\ba{\begin{array}}
\def\ea{\end{array}}

\def\tr{\mbox{Tr}}
\begin{document}
\maketitle
\begin{abstract} The law of track formation in cloud chambers is derived
from the Liouville equation with a simple Lindblad's type generator that
describes coupling between a quantum particle and a classical, continuous,
medium of two--state detectors. Piecewise deterministic random process
(PDP) corresponding to the Liouville equation is derived. The process
consists of pairs $(classical\, event, quantum\, jump)$, interspersed with
random periods of continuous (in general, non--linear)
Schr\"odinger's--type evolution. The classical events are flips of the
detectors -- they account for tracks. Quantum jumps are shown, in the
simplest, homogeneous case, to be identical to those in the early
spontaneous localization model of Ghirardi, Rimini and Weber (GRW). The
methods and results of the present paper allow for an elementary derivation
and numerical simulation of particle track formation and provide an
additional perspective on GRW's proposal. \end{abstract}
\newpage
\section{Introduction}
Inspired by John Bell's challenging call for an {\sl exact}\, formulation
of quantum measurement theory \cite{bel1,bel2}, Ph.Blanchard and the
present author proposed a model of quantum measurement based on completely
positive (CP) semigroup coupling between a quantum system and a classical
one \cite{bla1}. The main advantages of this proposal emerged only after
the publication of \cite{bla1}. In the following series of papers
\cite{bla2} -- \cite{bla5} the method of Ref. \cite{bla1} was successfully
applied to several model physical situations, including Zeno's effect,
Stern-Gerlach -- type coupling, particle position detector, and SQUID--tank
system. In all those cases the coupling was shown to lead to a piecewise --
deterministic random process (PDP) describing time series of experimentally
observed {\sl events}\,. Moreover, in Ref. \cite{jad1} models that deal
with simultaneous measurement of several non--commuting observables were
described, and it was suggested that the question of determining an unknown
state of the quantum system should be answered using the proposed exact
definition of a measurement. However, the obvious and crucial test of any
quantum measurement theory, namely that of finding the laws governing track
formation in cloud chambers and on photographic plates was, until recently,
missing. The reason for this was partly of a technical character, namely in
all of these previous applications the classical system was either discrete
or finite -- dimensional, otherwise technical difficulties mounted. In the
present paper we will show how these difficulties can be overcomed owing to
the discrete Poisson nature of the PDP.

Technically, the paper is concerned with a non -- relativistic quantum
particle coupled to a classical medium of two--state particle detectors.
The medium is characterized by a family of "sensitivity" functions
$g_a(x)$, where $g_a$ can be thought of as a Gaussian--like function
centered at $a$. \footnote{If there is no detector at $a$, we put
$g_a(x)\equiv0$. Thus our model covers also the case of a discrete, finite
or infinite, number of detectors.} The configuration space of the classical
system is, in general, infinite--dimensional. In $\$ 3$ we will write down
the simplest possible Liouville equation (Eqs.
(\ref{eq:liou})--(\ref{eq:va})) corresponding to the intuitive idea that
presence of the particle at some point $a$ causes flip in the detector
located at that point. The functions $g_a$ are used to describe the spatial
sensitivity, and also the response time, of the detectors. The quantum
Hamiltonian is allowed to depend on the actual configuration of the medium
(although in most applications such a dependence can be neglected). We
denote by $H_\Gamma$ the Hamiltonian corresponding to detectors flipped at
the points of a set $\Gamma$. The main result of the present paper is the
derivation of a PDP corresponding to this coupling. The PDP, derived in $\$
4$, can be summarized as follows.

\medskip
{\sl Suppose one starts, at time $t_0$, with all detectors in the "off"
state, and with the quantum object described by a wave function $\psi
_0=\psi_{t_0}$. Then $\psi$ evolves continuously according to the modified
Schr\"odinger evolution
$$\psi_{\Gamma;t}=\exp\left(-iH_\Gamma t-{{\Lambda t}\over 2}\right)\psi_0
/{\Vert \exp\left(-iH_\Gamma t-{{\Lambda t}\over 2}\right)\psi_0\Vert},$$
with $\Gamma=\{\emptyset \}$ and with $\Lambda$ defined by $\Lambda
(x)=\int g_a(x)^2 da$, until a jump occurs at a random time $t_1$, at which
time the wave function is, say, $\psi_{t_1}$. The jump consists of a pair:
{\sl (classical event,quantum jump)}\,. The classical part is a flip of the
detector state at a random point of space, say at $a_1$. It happens at a
point $a$, with the probability density $p(a)$ given by  $p(a)=\Vert g_a
\psi_{t_1}\Vert^2/\lambda(\psi_{t_1})$,
where the rate function $\lambda$ is given by
$\lambda(\psi)=(\psi,\Lambda\psi)$. The quantum part of the jump is jump of
the Hilbert space vector $\psi_{t_1}$ to the new state
$\psi_1=g_{a}\psi_{t_1}/{\Vert g_{a}\psi_{t_1}\Vert}$. After the jump the
process starts again with a continuous time evolution as before, but now
with $\Gamma=\{a_1\}$. After $n$ events, that happened at the points
$a_1,\ldots ,a_n$, one puts $\Gamma=\{a_1,\ldots ,a_n\}$. The random times
of jumps are regulated by an inhomogeneous Poisson process: the probability
$P(t,t+\triangle t)$ for the first jump to occur in the time interval
$(t,t+\triangle t)$ is computed from the formula
$P(t,t+\triangle t)=1-\exp \left(-\int_t^{t+\triangle t}
\lambda(\psi_s)\right)ds\approx \lambda(\psi_t)\triangle t$.}
\medskip

Our model admits an interesting special case - that of a passive,
homogeneous medium. If the medium is passive, i.e. if the quantum
Hamiltonian does not depend on the actual state of the medium, and if it is
homogeneous, then the description simplifies: the quantum process
separates, the jump rate is constant, and one gets "spontaneous
wave--packet reductions" of Ghirardi--Rimini--Weber (cf. e.g. \cite{ghi1}
and references therein). In general, however, the process of formation of a
track has a non--constant rate, and the dependence of the rate of jumps on
the state of the quantum system given by the present model is essential and
experimentally verifiable \footnote{This is one of the important
differences between our approach and other ones, where dependence of the
timing of wave packet reductions on the actual state of the quantum system
could not be derived - cf. e.g.\cite{cav1} and references therein.}. We
believe that the proposed model of the particle track formation is the
simplest one that gives intuitively expected results. It can be used for
numerical simulation of particle track formation for different Hamiltonians
and for different geometric configurations. It should be, in particular,
interesting to analyze numerically the influence of particle detectors on
sharpness of the fringe pattern in interferometry experiments.\\
{}From a philosophical point of view, it is worth noting that in the present
paper we deliberately avoid the concepts of an "observer". Our model aims
at being totally objective. A philosophical summary of our results can be
formulated as follows: Quantum Theory, once invented by human minds and
ones asked questions that are of interest for human beings, needs not
"minds" or "observers" any more. What it needs is lot of computing power
and effective random number generators, rather than "observers". The
fundamental question, to which we do not know answer yet, can be thus
formulated as follows: can random number generators be avoided and replaced
by deterministic algorithms of simple and clear meaning?

\section{Events and Quantum Measurements}
In this paragraph we will briefly describe the main ideas that influenced
our way of looking at the quantum mechanical measurement problem, and that
finally led to the simple cloud chamber model of this paper.

The crucial concept of our approach to quantum measurements is that of an
"event". The importance of this concept, and the intrinsic incapability of
quantum theory to deal with it, have been stressed by several authors. In
1958 E. Schr\"odinger wrote \cite{sch1}:
\begin{quotation} \lq It is usually believed that the current orthodox
theory actually accounts for the "nice linear traces" produced in the
Wilson chamber etc. I think this is a mistake, it does not.\rq
\end{quotation}
H.P. Stapp stressed the role of "events" in the "world process" (Refs.
\cite{stap2,stap3}, cf. also the entry "events" in the Index of
\cite{stap4}). G. F. Chew used Stapp's ideas on soft--photon
creation--annihilation processes (cf. \cite{stap5}) and proposed the term
"explicate order", complementing Bohm's "implicate" quantum order, to
denote the world process of "gentle" creation--annihilation events
\cite{chew1}. R. Haag emphasized \cite{haag1} that "an event in quantum
physics is discrete and irreversible" and that "we must assume that the
arrow of time is encoded in the fundamental laws ...". In \cite{haag2} he
went on to suggest that "transformation of possibilities into facts must be
an essential ingredient which must be included in the fundamental
formulation of the theory".\\
In \cite{bel1,bel2} J.S. Bell reprimanded the misleading use of the term
"measurement" in quantum theory. He opted for banning this word from our
quantum vocabulary, together with other vague terms such as "macroscopic",
"microscopic", "observable" and several others. He suggested to replace the
term "measurement" by that of "experiment", and also not to speak of
"observables" (the things that seem to call for an "observer") but to
introduce instead the concept of "beables" - the things that objectively
"happen--to--be (or not--to--be)". \footnote{Calling observables
"observables" can be, however, justified in the framework of an objective
theory of experiments. We plan to discuss this subject elsewhere .}

On the technical side, S. Machida and M. Namiki \cite{mach1} proposed a way
of describing measurements in quantum mechanics that inspired H. Araki
\cite{ara1,ara2} to formulate his continuous superselection rule model of
classical measuring apparatus in quantum mechanics. In the Araki's model
infinite time was however needed for an "event" (change of the classical
pointer position) to occur.

In a series of papers E.C.G. Sudarshan et al.  investigated possibility
 of solving the measurement problem via a unitary, Hamiltonian
coupling between a quantum and a classical system (cf. \cite{su1} and
references therein).

N.P. Landsman \cite{lan1} and M. Ozawa \cite{oza1} gave quite general
("no--go") arguments that stressed impossibility of coupling of classical
and quantum degrees of freedom via a unitary, finite--time dynamics.
\footnote{A short no--go argument can be found in Ref. \cite{jad1}.}

On the other hand many authors were using "dynamical semigroups" -- {\sl
non--unitary} dissipative time--evolutions that described an effective
dynamics of quantum systems coupled to other quantum systems or to external
"reservoirs" or "environment". V. Gorini et al. \cite{gor1} and Lindblad
\cite{lin1} derived a general form of generators of norm--continuous
semigroups of completely positive maps of the operator algebra of a Hilbert
space. \footnote{It was later extended by E. Christensen and D. Evans
\cite{chri1} to cover the case of more general operator algebras, including
the case that is most interesting for us - that of a non--trivial centre.}
Such semigroups were widely applied to many kinds of "master equations" of
statistical physics, while Ghirardi, Rimini and Weber \cite{ghi1} proposed
to use a particular Lindblad--type generator for describing a "spontaneous
reduction process" for a single quantum particle. The GRW model
incorporated "quantum jumps" that occurred in finite (Poisson distributed)
times, but it did not account for the (classical) "events". Although it was
clear to the experts that using dissipative semigroups instead of a unitary
dynamics allowed to go around the no--go theorems, it is only in
\cite{bla1} that simple methods of construction of dissipative generators
were found that led to measurement--like couplings of quantum and classical
degrees of freedom. Later on, in Refs. \cite{bla2}--\cite{jad1}, using the
results of M. H. A. Davis (see \cite{davmha1,davmha2}), a
piecewise--deterministic random process (PDP) on the space of pure states
of the total (classical+quantum) system was associated with the Liouville
equation. While the Liouville equation describes continuous time--evolution
of density matrices, that is of statistical states that concern {\sl
ensembles}, the associated piecewise--deterministic random process contains
apparently more useful information: it can be used to simulate real--time
behaviour of {\sl individual}\, systems in measurement--like situations.

\section{The Cloud Chamber Model}
Our aim is to explain the "nice linear tracks" that quantum particles leave
on photographs and in cloud chambers. These tracks are indeed hard to
explain if one assumes that there are no particles and no events -- only
Schr\"odinger's waves. Schr\"odinger himself was perplexed and not quite
sure which way to take.

Physically, a photographic plate or a cloud chamber is a highly complex
many-particle system. Physiologically, it appears to exhibits a complex,
irreversible dynamics to an external living observer. Many factors
participate in the result -- including the mediation of photons in the
final act of perception. However, it seems to us that the detailed internal
structure of local particle detectors, and also the details of the
perception process, would it be human or animal, are totally irrelevant for
the phenomenon itself. What {\sl is} relevant, it is the response of the
detectors to the quantum particle, and their back reaction on it. We put
forward conjecture that it is sufficient to assume that we have to do with
a system of {\sl classical}\, two--state detectors that can change their
state when a particle passes nearby. Although the real cloud chamber have a
finite number of sensitive centers, it proves to be no more difficult to
deal with a more general, continuous model - the extra bonus being that we
cover this way the GRW model as well.

There is a formal detail in the model below that deserves to be mentioned:
our model is more {\sl reversible} than any real cloud chamber. Namely, we
allow for a local detector to change its state {\sl back}, when it
registers the particle for the second time, and so on. This makes the model
slightly easier to solve. \footnote{On the other hand, it is related to
the, so called, "detailed balance condition" that is often postulated in
statistical physics models - for a recent discussion cf. \cite{stree1} and
references therein.} The present model can be easily reformulated to cover
also the case of "only-one-flip" detectors. The final PDP proves to be the
same except that each detector can flip only ones.

The derivation of the model below is heuristic. Nevertheless it leads to a
well defined piecewise--deterministic random process that has a clear
physical meaning. We then show that for a passive, homogeneous, medium, the
effective time evolution of the quantum system itself happens to be also
Markovian - it is described by an effective CP semigroup that is identical
to that postulated by Ghirardi et al. \cite{ghi1}. This fact may suggest
another application of our model: instead of considering it as an
approximate model of a real, discrete and finite, cloud chamber, we may
consider it as an exact model of some, perhaps yet unknown,
space--structure that is participating in a universal process of wave
packet reductions. The actual physical interpretation may depend on the
values of parameters that enter the model. There will be, essentially, two
free parameters: a coupling constant $\lambda$, of physical dimension
$t^{-1}$, that will regulate the expected time rate of jumps, and a
normalized Gaussian function whose width determines space sensitivity of
the detectors. In fact, aiming at a wider applicability of our model, we
will allow for non--constant rates of jumps, and for more general, not
necessarily Gaussian, sensitivity functions. Clearly, presence of arbitrary
functions that are external to the model, makes it to look like a
phenomenological rather than as a fundamental description -- unless these
functions are derived from geometrical and probabilistic considerations.

We proceed to describe our model in mathematical terms. The description
will be brief and will never go beyond elementary mathematical concepts.
Special mathematical terms, when they occur, are used only in an informal
way and can be skipped by a reader who is mainly interested in the main
ideas and results.\\ Let $E$ denote the physical space, we take for
definiteness $E={\bf R}^n$, although it is straightforward to $E$ to be a
homogeneous space or an arbitrary Riemannian manifold. We consider the
space $E$ filled up with a continuous medium which can be, at each point
$a\in E$, in one of its two states: "on--state", represented by
${1\choose0}$, or "off--state", represented by ${0\choose1}$. We would like
to consider the set of all possible states of the medium. This is however
enormously big a set, because states of the medium are, in our case, in
one--to--one correspondence with its configurations, that is with subsets
of $E$. Indeed, to each state of the medium we can uniquely associate the
set of all points that are "on". Thus the set of all states of the medium
is isomorphic to $2^E$. Fortunately we can restrict our attention to much
smaller classes of subsets of $E$. Let us introduce equivalence relation
$"\sim "$ in $2^E$, with equivalence classes consisting of subsets of $E$
that differ one from another by at most finite number of elements. Denoting
by ${\scriptstyle\triangle}$ the set--theoretical symmetric difference
operation, we have: $\Gamma\sim \Gamma '$ iff
$\Gamma{\scriptstyle\triangle} \Gamma '$ is a finite set. It will be
sufficient for us to choose some "ground state" and to take its equivalence
class, that is the set of these configurations that differ from the
"vacuum" in at most finite number of points. For convenience we will take
for the ground state the state of "all off", represented by the empty
subset $\emptyset \in 2^E$. Its equivalence class ${\cal S} = [ \emptyset
]$ consists of those states of the medium that are everywhere "off" except
in a finite number of points, i.e. the class of all {\sl } finite subsets
of $E$.
\vskip10pt
\noindent{\bf Remark.}
The fact that we can restrict ourselves to the above class ${\cal S}$ of
sets, instead of dealing with whole of $2^E$, is not evident by itself. It
will be justified only a posteriori, when we will see that the "events",
that will appear in the piecewise-deterministic random process which we
will construct later on, consist of "flipping" a state of the medium in
single (randomly chosen according to appropriate probability distribution)
points of $E$, and that with probability one there is a finite number of
events in any finite interval of time.
\vskip10pt
We can endow ${\cal S}$ with a topology and with a measurable structure as
follows: first of all we observe that ${\cal S}$ is a disjoint union of
subsets ${\cal S}_i , \, i=0,1,\ldots $, where ${\cal S}_i$ consists of
those states that differ from the ground state at exactly $i$ points of
$E$. But then, ${\cal S}_i$ is isomorphic to the $i-$th Cartesian power of
$E$, with coinciding points extracted, and divided by the action of the
permutation group in $i$ elements. It follows in particular that ${\cal S}$
has a power of the continuum.

Statistical states of the classical system are probability measures on
${\cal S}$. They are represented by sequences $\{ \mu_i \}$, where $ \mu_i$
is a measure on ${\cal S}_i$, and $\sum_{i=0}^{\infty }\mu ({\cal S}_i) =1$.

Let ${\cal H}_q= {\sl L}^2({\bf R}^n,d^n x)$ be the Hilbert space that is
used for description of the quantum system coupled to our classical medium.
We denote by ${\cal B} ({\cal H}_q )$ the algebra of bounded linear
operators on ${\cal H}_q$. Its elements are "observables" of the quantum
system. Statistical states of the quantum system are normalized (by $\tr
(\rho )=1$) positive trace class operators on ${\cal H}_q$. Then
statistical states of the total, classical {\sl plus}\, quantum, system are
described by measures $\rho $ on ${\cal S}$ with values in positive, trace
class, operators on ${\cal H}_q$, with $\sum_{i=0}^{\infty }Tr ( \rho
({\cal S}_i )) =1$. A natural candidate for the algebra ${\cal A}_{tot}$ of
observables of the total system is the algebra of continuous, bounded
functions on ${\cal S}$ with values in ${\cal B} ({\cal H}_q )$. Thus,
${\cal A}_{tot}$ is the direct sum of algebras ${\cal A}_i$, where ${\cal
A}_i$ is the algebra of continuous, bounded, ${\cal B} ({\cal
H}_q)$--valued functions on ${\cal S}_i$. As our main aim is to derive the
PDP rather than to prove the existence of CP semigroup -- we will apply,
from now on, a heuristic notation. Thus, a state of the total system will
be represented by a family $\{ \rho_\Gamma \}_{\Gamma\in {\cal S}}$, with
\lq$\sum_\Gamma$\rq $\tr(\rho_\Gamma)=1$.

To have some definite example in mind, in what follows we will take for the
quantum system a particle of mass $m$ moving in $E={\bf R}^n$ according to
the dynamics described by the quantum Hamiltonian \be H_\Gamma=
-{\hbar^2\over 2m}\left({\bf\nabla}_x-{e\over {\hbar c}}{\bf A}_
\Gamma\right)^2+V_\Gamma(x). \ee We thus allow quantum Hamiltonian to
depend on the actual state of the medium.
\vskip10pt
{\noindent\bf Remark}
We could allow $H$ to depend explicitly on time - then the semigroup
property would be lost, but PDP would be described in exactly the same way
as in the prsesnt model, except that the exponential in the formula
(\ref{eq:schr}) would have to be time-ordered. Generalization to the case
of quantum particle moving on a manifold and acted upon by gravitational
and electromagnetic forces is straightforward. A more general treatment,
including Bose or Fermi multiparticle case, will appear elsewhere
\cite{jad2}. The idea will not change also in such a case.
\vskip10pt
We proceed now to describe the coupling that corresponds to the following
intuitive picture: {\sl the medium consists of detectors that can change
their state if the particle approaches them sufficiently close.}\, The
sensitivity of detectors as well as their relaxation time are described by
real, non--negative functions $g_a(x)$, where the variable $a$ describes
the position of the detector. We can think of $g_a$ as a hat - like
function with its center at $x=a$. We introduce then the non--negative
function $\Lambda (x)$ defined by \be \Lambda (x)=\int_E g_a(x)^2 da ,
\label{eq:lam} \ee for all $x\in E$. Here $da$ denotes the Lebesgue
measure, but if we want to describe a discrete, rather than a continuous,
case, then the integral above should be replaced by a sum. By the abuse of
notation we will denote by the letter $\Lambda$ the operator of
multiplication by the function $\Lambda (x)$, acting on the Hilbert space
${\sl L}^2({\bf R}^n,d^n x)$.\\ Each state $\rho$ of the total system can
be, formally, written as: \be \rho = \sum_{\Gamma\in{\cal S}} {\rho}_\Gamma
\otimes {\epsilon}_\Gamma , \ee where, for $\Gamma\in{\cal S}$, \be
\epsilon_\Gamma =\prod \otimes_{a\in E}\pmatrix{{\chi}_\Gamma (a) & 0\cr
0&1-{\chi}_\Gamma (a))\cr}, \ee and where ${\chi}_\Gamma$ stands for the
characteristic function of $\Gamma$.
\vskip10pt
{\noindent\bf Remark}
The last statement requires some care. It is also not quite trivial. For a
finite number of detectors it is not too difficult to see. We are using
above the notation introduced by J. von Neumann in his theory of continuous
tensor products. To give to the above expressions a precise mathematical
meaning, we would have to invoke a part of this theory. (For a more modern
account cf. \cite{gui1} and references therein.) That tool is however not
necessary for the present, heuristic, purpose. More complete mathematical
treatment will be given elsewhere.
\vskip10pt
To define the coupling between the particle and the medium, we will apply
the ideas introduced in \cite{bla1,bla2}. Namely, we will write the
Liouville time evolution equation for the statistical state of the total
system as \be {\dot\rho}=-i[ H,\rho] + {\cal L}(\rho ) , \label{eq:liou}
\ee where ${\cal L}$ is a Lindblad--type generator that provides
dissipative coupling. \vskip10pt
{\noindent\bf Remark:} In
the present model we will neglect a possible free dynamics of the medium.
\vskip10pt
For ${\cal L}$ we take the simplest possible coupling: \be {\cal L}(\rho
)=\int da (V_a \rho V_a -{1\over 2}\{V_a^2,\rho\}) \label{eq:lin} \ee where
\be V_a=g_a\otimes \tau_a , \label{eq:va} \ee $g_a$ being the
multiplication operator by the function $g_a(x)$, and $\tau_a$ denoting the
"flip" of the detector at the point $a$: \be \tau_a=\prod\otimes_b u_b ,
\ee where \be u_b=\pmatrix{1&0\cr 0&1\cr} \ee for $b\ne a$, while \be
u_a=\pmatrix{0&1\cr 1&0\cr}. \ee Because $\tau_a^2=Id$, our evolution
equation reads now: \be {\dot\rho }=-i[ H, \rho]+\int da V_a\rho V_a -
{1\over 2}\{\Lambda, \rho\}, \ee where $\Lambda$ in the anticommutator is
understood as a multiplication operator by the function $\Lambda (x)$.\\
Let us denote by $a(\Gamma)$ the set representing the state $\Gamma$ with
the flipped $a$: \be a(\Gamma)=\Gamma{\scriptstyle\triangle}\{a\}. \ee
Then, using change of summation variable $\Gamma\rightarrow
\Gamma'=a(\Gamma)$, and also using the fact that $a(a(\Gamma))=\Gamma$ --
i.e. that the second flip cancels the first, we obtain \be V_a \rho V_a =
\sum_{\Gamma\in {\cal S}} g_a \rho_\Gamma g_a \otimes\epsilon_{a(\Gamma)} =
\sum_{\Gamma\in{\cal S}} g_a\rho_{a(\Gamma)} g_a\otimes\epsilon_\Gamma \ee
so, that we can write: \be {\dot \rho}_\Gamma=-i[H_\Gamma,\rho_\Gamma]+\int
da\, g_a \rho_{a(\Gamma)}g_a - {1\over 2}\{\Lambda\, ,\rho_\Gamma
\}.\label{eq:main} \ee The equation (\ref{eq:main}) fundamental. It
describes time evolution of the family $\{\rho_\Gamma\}$, where $\Gamma$
runs over all finite subsets of $E$. All the relevant statistical
information about the quantum particle and the classical medium can be
derived from this equation. In the next paragraph we will derive the
piecewise deterministic random process that is compatible with Eq.
(\ref{eq:main}) and that concerns histories of individual coupled systems.
Before however going to this, let us see that in the passive, homogeneous,
case we can obtain effective time evolution for the quantum particle alone.
If the medium is passive and homogeneous, then the Hamiltonian does not
depend on the actual state of the medium: $H_\Gamma\equiv H$. Moreover, for
symmetry reasons $\Lambda$ must be a constant: $\Lambda(x)\equiv\lambda$.
For instance this happens if we take for $g_a$ the Gaussian functions: \be
g_a(x)=\lambda^{1/2}\left({\alpha\over\pi}\right)^{n\over 2}
\exp\left({-\alpha (x-a)^2}\right). \ee The effective state of the quantum
particle is determined by tracing over the classical configurations: \be
{\hat\rho}=\sum_{\Gamma\in{\cal S}} \rho_\Gamma. \ee To sum up the Eq.
(\ref{eq:main}) over $\Gamma$ we note that, for each $a\in E$,
$a:\Gamma\mapsto a(\Gamma)$ is a one--to--one map of ${\cal S}$ onto
itself, this owing to the fact that $2^E$ is a group under the
symmetric--difference operation, and that ${\cal S}$ is a subgroup.\\ Thus
we have $\sum_{\Gamma\in{\cal S}}\rho_{a(\Gamma)}= \sum_{\Gamma\in{\cal
S}}\rho_\Gamma={\hat\rho}.$ It follows that the time derivative of
${\hat\rho}$ depends, for our particular choice of the coupling, only on
${\hat\rho}$ and not on the full hierarchy of $\rho_\Gamma$'s; we have: \be
{\dot {\hat\rho}} = -i[H,{\hat\rho}]+\int da\, g_a{\hat\rho} g_a -\lambda\,
{\hat\rho },\ee which is exactly of the type discussed by Ghirardi, Rimini
and Weber (cf. Ref. \cite{ghi1}).
\section{The Piecewise Deterministic Process}
\subsection{Definition of  PDP  and its infinitesimal generator}
In his monographs \cite{davmha1,davmha2} dealing with stochastic control
and optimization M. H. A. Davis, having in mind mainly queuing and
insurance models, described a special class of piecewise deterministic
processes that was later found to fit perfectly the needs of quantum
measurement theory. Even if for the present model we will have to extend
slightly the original Davis' framework, and to work with jumps between
continuously parameterized states and not between discrete manifolds, we
will describe briefly the discrete case and we leave the problem of a
rigorous formulation of its evident extension to continuous families
aside.

Let $\iota$ be an index running over a finite or countable set $J$.
Consider functions $f(\xi ,\iota)$, where for each $\iota$ the variable
$\xi $ is continuous and runs through some set $M$ \footnote{We will need
the case where also $\iota$ will be continuous running over $E$, while $M$
will coincide with the unit ball in the Hilbert space $L^2(E)$ (modulo the
phase).} Suppose we have a one--parameter semi-group of transformations
$\alpha_t$ acting on the space of such functions with the infinitesimal
generator ${\cal D}$ which is an integro--differential operator of the
following form: \be \ba{c} \left({\cal D}f\right)(\xi,\iota) =(Z_\iota
f)(\xi,\iota)\cr\cr + \lambda(\xi,\iota)\sum_{\iota'} \int_M
Q(\xi,\iota;d\xi',\iota') \left( f(\xi',\iota')-f(\xi,\iota)\right)
,\label{eq:gen} \ea \ee where $Z_\iota$ are vector fields that generate
one--parameter flows $\phi_\iota$ on $M$, $\lambda(\xi,\iota)$ are
non--negative functions, while $ Q(\xi,\iota;d\xi',\iota')$ are
(non--negative) transition measures - thus satisfying \be \sum_{\iota'}
\int_M Q(\xi,\iota;d\xi',\iota') =1 , \ee and also \be
\int_{\{\xi\}}Q(\xi,\iota;d\xi',\iota)=0 , \ee for all $\iota$ and $\xi\in
M$. We notice that by the very definition we have $Z_\iota(\xi)=
d\phi_\iota(\xi,t)/dt\, \vert_{t=0}$. Then, as it is shown in Refs.
\cite{davmha1,davmha2}, one can associate with this generator ${\cal D}$ a
piecewise--deterministic Markov process that is described as follows.

Suppose the process starts at some point $(\xi_0 ,\iota_0 )$. Then $\xi$
evolves continuously along the vector field $Z_\iota$,
$\xi_t=\phi_\iota(\xi_0,t)$, while $\iota_0$ remains constant until a jump
occurs at a certain random time $t_1$. The time of this jump is governed by
a (inhomogeneous) Poisson process with rate function
$\lambda(t)=\lambda(\xi_t,\iota_0)$. When jump occurs at $t=t_1$, then
$(\xi_{t_1},\iota_0)$ jumps to $(\xi',\iota)$ with probability density
$Q(\xi_{t_1},\iota_0;d\xi', \iota)$ and the process starts again.\\
\vskip10pt
{\bf Remark} Notice that the probability that the jump will occur between
$t$ and $t+dt$, provided it did not occurred yet is equal to
$1-\exp\left(-\int_t^{t+dt}\lambda(s)ds\right)\approx\lambda(t)dt$. This
justifies calling $\lambda$ the rate function.
\vskip10pt
Association of the random process with the semi--group $\alpha_t$ is
canonical and can be described as follows: first one goes from $\alpha_t$
that acts on functions $f(\xi,\iota)$ to its dual $\alpha^t$ acting on
measures. Then, choosing the Dirac measure $\delta_{\xi_0,\iota_0}$
concentrated at $(\xi_0,\iota_0)$ as the initial point $\mu_0$, we apply to
it $\alpha^t$ to get $\mu_t=\alpha^t(\mu_0)$. The resulting measure $\mu_t$
is then characterized by the fact that $d\mu(\xi,\iota)$ is equal to the
probability that the process starting at $t=0$ from $(\xi_0,\iota_0)$ will
end, at time $t$, at the point $(\xi,\iota)$.\\ A detailed and precise
description of the above correspondence should include specification of the
involved measure structures and domains of definition. We refer the reader
to Refs. \cite{davmha1,davmha2} for mathematical details.
\subsection{Derivation of the PDP for the cloud chamber model.}
We will now describe the most important fact about our cloud chamber
model: we will show that Eq. (\ref{eq:main}) describing the time
evolution of statistical states of the total system can be interpreted
in terms of a piecewise deterministic random process. That process has
then a transparent description in terms of pairs of
{\sl (classical event,quantum jump)}\, that are interspersed (in a random
way, according to an homogeneous Poisson point process law with rate
$\lambda$) with the periods of continuous, Schr\"odinger's type, time
evolution.

If we want to interpret the equation (\ref{eq:main}) in terms of a PDP on
pure states, then the first thing we have to do, is to rewrite
Eq.(\ref{eq:main}) as an equation for observables rather then states. After
doing so we will interpret observables as functions on pure states.\\ Given
a state $\rho=\{\rho_\Gamma:\Gamma\in{\sl S}\}$ and an observable
$A=\{A_\Gamma:\Gamma\in{\sl S}\}$, the expectation value of $A$ in $\rho$
is given by $<\rho,A>=\sum_\Gamma \tr (\rho_\Gamma A_\Gamma)$. Time
evolution of observables is then defined as dual to the time evolution of
states, so that we have $<\rho,\dot A>=<{\dot \rho},A>$. By substituting
the equation (\ref{eq:main}) for ${\dot \rho}$, we easily find that, in our
case, observables evolve according to the law that is almost identical to
that for states, except that there is change of sign in front of the
commutator: \be {\dot A}_\Gamma=i[H_\Gamma,A_\Gamma]+\int da\, g_a
A_{a(\Gamma)}g_a - {1\over2}\{\Lambda,A_\Gamma\} .\label{eq:maina} \ee Each
observable $A$ (of the total system) can be interpreted as a function $f_A$
on pure states (of the total system): \be
f_A(\psi,\Gamma)\equiv(\psi,A_\Gamma\psi),\qquad \psi\in {\cal H}_q,\,
\Gamma\in {\cal S}. \ee We can now sandwich the Eq.(\ref{eq:maina}) between
two $\psi$ vectors to see if we can interpret this equation in terms of
time evolution of functions on pure states. We get \be \ba{cc} {\dot
f}_A(\psi,\Gamma)\equiv f_{\dot A}(\psi,\Gamma)\\
=(\psi,i[H_\Gamma,A_\Gamma]\psi) +(\psi,\int da\, g_a A_{a(\Gamma)}g_a\,
\psi)-{1\over 2}(\psi,\{\Lambda, A_\Gamma \} \psi ). \ea\label{eq:mainb}
\ee The first term on the rhs of Eq. (\ref{eq:mainb}) can be written also
as $(Z_H f_A)(\psi,\Gamma)$, where $Z_H$ is the vector field of the
Hamiltonian evolution of pure states \be (Z_H f) (\psi,\Gamma)\doteq
{d\over dt}f\left (e^{-iH_{\Gamma} t} \psi,\Gamma \right)\bigg\vert_{t=0} \
.\label{eq:ham} \ee The second term can be rewritten as: \be \ba{c}
(\psi,\int da\, g_a A_{a(\Gamma)}g_a\, \psi) =\int da\,
(g_a\psi,A_{a(\Gamma)}g_a\psi)\cr\cr =(\psi,\Lambda\psi)\int da\, {\Vert
g_a\psi\Vert^2 \over {(\psi,\Lambda\psi )}} f_A({g_a\psi\over {\Vert
g_a\psi\Vert}},a(\Gamma)). \ea\label{eq:second} \ee Finally, the third term
of the Eq. (\ref{eq:mainb}), rewritten in terms of the functions $f_A$,
gives rise to two terms: \be \ba{c} -{1\over
2}(\psi,\{\Lambda,A_\Gamma\}\psi )= {d\over dt}\left( \Vert \exp\left(
-{\Lambda\over 2}t\right)\psi\Vert^2 f_A\left({\exp\left( -{\Lambda\over
2}t\right)\psi\over \Vert \exp\left( -{\Lambda\over
2}t\right)\psi\Vert},\Gamma\right)\right) \vert_{t=0}\cr\cr =
-(\psi,\Lambda\, \psi ) + {d\over dt}\left( f_A\left({\exp\left(
-{\Lambda\over 2}t\right)\psi\over \Vert \exp\left( -{\Lambda\over
2}t\right)\psi\Vert},\Gamma\right) \right)\vert_{t=0}. \ea \ee Let us
introduce the second vector field $Z_D$ corresponding to the non--linear
evolution: \be (Z_D f) (\psi,\Gamma)\doteq {d\over dt}f\left({\exp\left(
-{\Lambda\over 2}t \right)\psi\over \Vert \exp\left( -{\Lambda\over
2}t\right)\psi\Vert},\Gamma\right)\vert_{t=0}. \ee We now see that we can
write the evolution equation for the functions $f_A$ in the form required
by Eq.(\ref{eq:gen}) provided we introduce:\\ the rate function: \be
\lambda(\psi)=(\psi,\Lambda\psi), \label{eq:lam2}\ee the vector field: \be
Z=Z_H+Z_D ,\ee and the transition measure
$Q(\psi,\Gamma;\psi',\Gamma')d\psi'd\Gamma'$ that vanishes except for \be
Q(\psi,\Gamma;\psi',a(\Gamma))={\Vert g_a \psi \Vert^2\over
{\lambda(\psi)}} \delta (\psi'- {g_a\psi\over{\Vert g_a\psi\Vert}})d\psi',
\ee where $\delta(\psi'-\psi)d\psi'$ is a symbolic expression for the Dirac
measure concentrated at $\psi$.\\ It is easy to see that the vector field
$Z=Z_H+Z_D$ exponentiates to: \be \left(\exp
(Zt)f\right)(\psi_0,\Gamma)=f\left(\psi_{\Gamma;t},\Gamma\right) \ee where
$\psi_{\Gamma;t}$ is given by \be \psi_{\Gamma;t}={\exp\left(-iH_\Gamma
t-{{\Lambda t}\over 2}\right)\psi_0 \over{\Vert \exp\left(-iH_\Gamma
t-{{\Lambda t}\over 2}\right)\psi_0\Vert}}.\label{eq:schr}\ee
Thus $\psi_{\Gamma;t}$ can be thought of as a solution of a non--linear,
non--Hermitian Schr\"odinger equation.

We now describe the piecewise deterministic process on pure states of the
total system that is associated with these data. Starting with the quantum
system described by an initial wave packet $\psi\in L^2(E)$, and with the
initial "all off" state of the medium, $\psi$ develops according to the
equation (\ref{eq:schr} until a jump occurs at random time $t_1$, at which
time the wave packet is $\psi_{t_1}$. The time $t_1$ of the jump is
governed by the inhomogeneous Poisson process that is characterized by the
probability $P(t,t+\triangle t)$ for the jump to occur in the time interval
$(t,t+\triangle t)$, provided it did not occurred yet. It is given by the
formula
\be
P(t,t+\triangle t)=1-\exp \left(-\int_t^{t+\triangle t}
\lambda(\psi_s)\right)ds\approx \lambda(\psi_t)\triangle t .
\label{eq:rt}
\ee
The jump consists of a pair {\sl (classical event,quantum jump)}\,. The
classical event is a flip of the detector at a random point $a\in E$. It
happens at $a$ with probability density
\be
p(a)={\Vert g_a \psi_{t_1}\Vert^2\over\lambda(\psi_{t_1})}.
\label{eq:pa}
\ee
 When the classical detector flips at some point $a=a_1$, then the quantum
states jumps from its actual state $\psi_{t_1}$ to the new state $\psi_1$
given by
\be
\psi_1={g_{a_1}\psi_{t_1}\over{\Vert g_{a_1}\psi_{t_1}\Vert}}
\ee
and the process starts again.\\
It is worth noting that, for simple Gaussian packets, and for a free
evolution, the most probable place for a flip to occur is at the maximum of
the actual wave--function. That explains linear tracks. For more
complicated geometries and dynamics -- numerical computation is necessary,
at least until simple general laws are found that are based on PDP.

\section{Summary and Conclusions}
We have seen that a simple coupling between quantum particle and classical
continuous medium of two--state detectors leads to a piecewise
deterministic random process that accounts for track formation in cloud
chambers and photographic plates. For a passive, homogeneous medium the
process is essentially identical to the spontaneous localization GRW model
of Ref. \cite{ghi1}. In particular all the theoretical and numerical
analysis that has been done for GRW models applies also in this case.\\ As
mentioned in the Introduction, to simulate track formations only random
number generators and computing power is necessary. Our model does not
involve observers and minds. This does not mean that we do not appreciate
the importance of the mind--body problem. In our opinion understanding the
problems of minds needs also quantum theory, and perhaps even more -- that
is still beyond the horizon of the present-day physics. But our model
indicates that quantum theory does not need human minds. Quantum theory
should be formulated in a way that involves neither observers nor minds -
at least not more than any other branch of physics. Our model can be
considered as a step in this direction. It can rightly be criticized as
being too phenomenological to satisfy us wholly. But, provided it correctly
accounts for experimental results, it can give a valuable new insight into
the quantum duality of potential and actual, of waves and particles, and of
determined and random.
\newpage
\noindent
{\bf Acknowledgments}
The necessity of formulating and of investigating a cloud chamber model, as
well as the rough idea of how to do it, was the subject of many discussion
with Prof. Ph.Blanchard. I thank him also for reading the manuscript. The
work was mainly done while the author stayed at RIMS, Kyoto. Thanks are due
to JSPS for financial support and to Prof. H. Araki for his kind
hospitality and for encouraging and useful discussions. Thanks are also due
to Profs. H. Ezawa, M. Namiki and I. Ojima for discussions and for their
criticism. I would also like to thank Prof. M. Ozawa for his criticism
concerning CP semigroup couplings. I owe lot of thanks to Prof. H.P. Stapp
for useful correspondence during all of the period of writing this
paper.\\
The final writeup was done at the Erwin Schr\"odinger Institute, Wien. I
thank to Prof. H. Narnhofer for her hospitality and for reading the paper.
I also owe thanks to Prof. F. Benatti for clarifying and encouraging
correspondence.


\begin{thebibliography}{99}
\bibitem[1]{bel1} Bell, J. : "Against measurement", in {\sl Sixty-Two Years
of Uncertainty. Historical, Philosophical and Physical Inquiries into the
Foundations of Quantum Mechanics}, Proceedings of a NATO Advanced Study
Institute, August 5-15, Erice, Ed. Arthur I. Miller, NATO ASI Series B vol.
226 , Plenum Press, New York 1990
\bibitem[2]{bel2} Bell,  J. : "Towards an exact quantum mechanics",  in
{\sl Themes in Contemporary Physics II.  Essays in honor of Julian
Schwinger's 70th birthday},  Deser,  S. ,  and Finkelstein,  R. J.  Ed. ,
World Scientific,  Singapore 1989
\bibitem[3]{bla1} Blanchard, Ph. and Jadczyk, A. : "On the interaction
between classical and quantum systems", {\sl Phys. Lett. }{\bf A 175}
(1993), 157--164
\bibitem[4]{bla2} Blanchard, Ph. and Jadczyk, A. : "Strongly coupled
quantum and classical systems and Zeno's effect", {\sl Phys. Lett. }{\bf A
183} (1993), 272--276
\bibitem[5]{bla3} Blanchard,  Ph.  and Jadczyk,  A. : "Classical and quantum
intertwine",  in {\sl Proceedings of the Symposium on Foundations of
Modern Physics},  Cologne,  June 1993,  Ed.  P.  Mittelstaedt,  World
Scientific  (1994), hep--th 9309112
\bibitem[6]{bla4} Blanchard,  Ph.  and Jadczyk,  A. : "From quantum
probabilities to classical facts",  in {\sl Advances in Dynamical
Systems and Quantum Physics},  Capri,  May 1993,  Ed.  R.  Figari,  World
Scientific  (1994), hep--th 9311090
\bibitem[7]{bla5} Blanchard,  Ph.  and Jadczyk,  A. :  "How and
When Quantum Phenomena Become Real",  to appear in Proc.
Third Max Born Symp.  "Stochasticity and Quantum Chaos",
Sobotka,  Eds.  Z.  Haba et all. ,  Kluwer Publ.
\bibitem[8]{jad1} Jadczyk, A. : "Topics in Quantum Dynamics", Preprint CPT--
Marseille 94/P.3022, also BiBoS 635/5/94, hep--th 9406204
\bibitem[9]{ghi1} Ghirardi, G.C., Rimini, A. and Weber, T. : "An Attempt at
a Unified Description of Microscopic and Macroscopic Systems", in {\sl
Fundamental Aspects of Quantum Theory}, Proc. NATO Adv. Res. Workshop,
Como, Sept. 2--7, 1985, Eds. Vittorio Gorini and Alberto Frigerio, NATO ASI
Series B 144, Plenum Press, New York 1986, pp. 57--64
\bibitem[10]{cav1} Caves, C. M. and Milburn, G. J. : "Quantum mechanical
model for continuous position measurements", {\sl Phys. Rev.} {\bf A 36}
(1987) 5543--5555
\bibitem[11]{stap1} Stapp, H. P. : "The Integration of Mind into
Physics", Talk at the conference Fundamental Problems in Quantum
Theory, Univ. of Maryland at Baltimore, June 18--22, 1994,
Preprint LBL-35640, June 1994
\bibitem[12]{sch1} Schr\"odinger, E. : "Might perhaps Energy be a merely
Statistical Concept?", {\sl Nuovo Cimento} {\bf IX} (1958) 162--179
\bibitem[13]{stap2} Stapp, H. P. : "Bell's Theorem and World Process",
{\sl Nuovo Cimento} {\bf 29} (1975) 270--276
\bibitem[14]{stap3} Stapp, H. P. : "Theory of Reality", {\sl Found. Phys.}
{\bf 7} (1977) 313--323
\bibitem[15]{stap4} Stapp, H. P. : "Mind, Matter and Quantum
Mechanics", Springer Verlag, Berlin 1993
\bibitem[16]{stap5} Stapp, H. P. : "Solution of the Infrared Problem",
{\sl Phys. Rev. Lett.} {\bf 50} (1983) 467--469
\bibitem[17]{chew1} Chew, G. F. : "Gentle Quantum Events and the Source
of Explicate Order", {\sl Zygon} {\bf 20} (1985) 159--164
\bibitem[18]{haag1} Haag, R. : "Events, histories, irreversibility", in
{\sl Quantum Control and Measurement}, Proc. ISQM Satellite Workshop,
ARL Hitachi, August 28--29, 1992, Eds. H. Ezawa and Y. Murayama,
North Holland, Amsterdam 1985
\bibitem[19]{haag2} Haag, R. : "Fundamental Irreversibility and the Concept
of Events", {\sl Commun. Math. Phys.}{\bf 132} (1990) 245--251
\bibitem[20]{mach1} Machida, S. and Namiki, M. : "Theory of
Measurement in Quantum Mechanics", {\sl Progr. Theor. Phys.}
{\bf 63} (1980), 1457--1473 and 1833--1847
\bibitem[21]{ara1} Araki, H. : "A Remark on Machida--Namiki Theory
of Measurement", {\sl Progr. Theor. Phys.} {\bf 64} (1980) ,
719--730
\bibitem[22]{ara2} Araki, H. "A Continuous Superselection Rule as a Model
of Classical Measuring Apparatus in Quantum Mechanics", in {\sl Fundamental
Aspects of Quantum Theory}, Proc. NATO Adv. Res. Workshop, Como, Sept.
2--7, 1985, Eds. Vittorio Gorini and Alberto Frigerio, NATO ASI Series B
144, Plenum Press, New York 1986, pp. 23--33
\bibitem[23]{su1} Sudarshan, E.C.G. : "Measurement Theory", in {\sl
Foundations of Quantum Theory}, Santa Fe Workshop 27-31 May 1993, Ed.
Black, T.D. et al., World Scientific, Singapore 1992
\bibitem[24]{lan1} Landsman,  N. P. : "Algebraic theory of
superselection sectors and the measurement problem in quantum
mechanics", {\sl Int. J. Mod. Phys. }{\bf A6} (1991), 5349--5371
\bibitem[25]{oza1} Ozawa, M.: "Cat Paradox for $C^{\star}$--Dynamical
Systems", {\sl Progr. Theor. Phys. }{\bf 88} (1992), 1051--1064
\bibitem[26]{gor1} Gorini,  V. ,  Kossakowski,  A.  and Sudarshan,
E. C. G. : "Completely positive dynamical semigroups of N--level
systems",  {\sl J.  Math.  Phys. }{\bf 17} (1976), 821--825
\bibitem[27]{lin1} Lindblad, G. : "On the Generators of Quantum Mechanical
Semigroups", {\sl Comm. Math. Phys. }{\bf 48} (1976), 119--130
\bibitem[28]{chri1} Christensen,  E.  and Evans,  D. : "Cohomology of
operator algebras and quantum dynamical semigroups",  {\sl J.  London.
Math.  Soc. }  {\bf 20} (1978), 358--368
\bibitem[29]{davmha1} Davis, M. H. A. : {\sl Lectures on Stochastic Control
and Nonlinear Filtering}, Tata Institute of Fundamental Research, Springer
Verlag, Berlin 1984
\bibitem[30]{davmha2} Davis, M. H. A. : {\sl Markov models and optimization},
Monographs on Statistics and Applied Probability,  Chapman and Hall,
London 1993
\bibitem[31]{stree1} Majewski, W. A. and Streater, R. F. : "Detailed
Balance and Quantum Dynamical Semigroups", preprint Dept. Math., King's
College, London, 1994
\bibitem[32]{jad2} Jadczyk, A. : "On Quantum Jumps, Events and Spontaneous
Localization Models", ESI-Wien  Preprint (1994)
\bibitem[33]{gui1} Guichardet, M. A. :"Produits Tensoriels Infinis et
Repr\'esentations des Relations d'Anticommutation", {\sl Ann. sc. \'Ec.
Norm. Sup.}{\bf 83} (1966), 1--52
\end{thebibliography}
\end{document}